\documentclass[showpacs,preprintnumbers,superscriptaddress]{revtex4}
\usepackage{amssymb}
\usepackage[centertags]{amsmath}
\usepackage{tipa}
\usepackage{txfonts}
\usepackage{epsfig}
\usepackage{bm}

\makeatletter
  \newcommand\figcaption{\def\@captype{figure}\caption}
  \newcommand\tabcaption{\def\@captype{table}\caption}
\makeatother

\begin{document}
\title{The influence of net-quarks on the yields and rapidity
spectra of identified hadrons}
\author{Jun Song }
\affiliation{Department of Physics, Qufu Normal University,
Shandong 273165, People's Republic of China}

\author{Feng-lan Shao}
\affiliation{Department of Physics, Qufu Normal University,
Shandong 273165, People's Republic of China}

\author{Qu-bing Xie}
\affiliation{Department of Physics, Shandong University,
Shandong 250100, People's Republic of China}

\author{Yun-fei Wang }
\affiliation{Department of Physics, Qufu Normal University,
Shandong 273165, People's Republic of China}

\author{De-ming Wei }
\affiliation{Department of Physics, Qufu Normal University,
Shandong 273165, People's Republic of China}

\begin{abstract}
Within a quark combination model,  we study systematically the
yields and rapidity spectra of various hadrons in central Au+Au
collisions at $\sqrt{s_{NN}}= 200$ GeV. We find that considering the
difference in rapidity between net-quarks and newborn quarks, the
data of multiplicities, rapidity distributions for $\pi^{\pm}$,
$K^{\pm}$, $p(\overline{p})$ and, in particular the ratios of
charged antihadron to hadron as a function of rapidity, can be well
described. The effect of net-quarks on various hadrons is analysed,
and the rapidity distributions for $K^{0}_{s}$,
$\Lambda(\overline{\Lambda} )$, $\Sigma^{+}(
\overline{\Sigma}^{\,_-})$, $\mathrm{\Xi^{-}}$
($\mathrm{\overline{\Xi}^{\,_+}}$) and
$\mathrm{\Omega^{-}}\!(\mathrm{\overline{\Omega}}^{_+})$ are
predicted. We discuss the rapidity distribution of net-baryon, and
find that it reflects exactly the energy loss of colliding nuclei.
\end{abstract}


\maketitle

\section{Introduction}
The relativistic heavy ion collider (RHIC) at Brookhaven National
Lab (BNL) provides a unique environment to search for the quark
gluon plasma (QGP) predicted by lattice QCD calculations
\cite{Blum:1995}, and to study the properties of this matter at
extremely high energy densities. A huge number of data have been
accumulated and used to extract the information about the original
partonic system and its space-time evolution. A variety of
experimental facts from various aspects imply that the strongly
coupled QGP has been probably produced in central Au+Au collisions
at RHIC
\cite{Adams:2005dq,Gyulassy:2004zy,Jacobs:2004qv,Kolb:2003dz,
Braun-Munzinger:2003zd,Rischke:2003mt}. Due to the confinement
effects, one can only detect the hadrons freezed out from the
partonic system  rather than make direct detection of partons
produced in collisions. Therefore, one of the important
prerequisites for exploring the quark gluon plasma is the better
understanding of hadronization mechanism in nucleus-nucleus
collisions.

The quark combination picture is successful in describing many
features of multi-particle production in high energy collisions. The
parton coalescence and recombination models have explained many
highlights at RHIC, such as the high ratio of $p/\pi$ at
intermediate transverse momenta
\cite{Fries:2003,Greco:2003prl,Hwa:2003} and the quark number
scaling of hadron elliptic flows
\cite{Voloshin:2002wa,Molnar:2003ff,Lin:2002rw}. Our quark
combination model has been used to describe the charged particle
pseudorapidity densities \cite{shao:2006}, hadron multiplicity
ratios, $p_{T}$ spectra \cite{Shao:2004cn} and elliptic flows
\cite{Yao:2006fk} at midrapidity.

The hadronic yields and rapidity spectra and, particularly the
rapidity dependence of net-baryon rapidity densities and
antihadron-to-hadron ratios, are important observables in high
energy nucleus-nucleus collisions, from which a lot of information
about the hot and dense matter can be extracted
\cite{bearden:2004,Herrmann:1999,Satz:2000}. The BRAHMS
Collaboration have measured the ratios of antihadron-to-hadron and
net-baryon rapidity density ($dN_{(B-\overline{B}\,)}/dy$) as a
function of rapidity in central Au+Au collisions at
$\sqrt{s_{NN}}=200$ GeV \cite{Bearden:2003,bearden:2004}. The data
show that the ratios of $K^{-}/K^{+}$, $\overline{p}/p$ decrease
obviously with increasing rapidity and the
$dN_{(B-\overline{B}\,)}/dy$ exhibits  a dual-hump shape in the full
rapidity region. It can be explained by a positive and rapidity
dependent baryon chemical potential in statistic thermal
model\cite{Braun-Munzinger:2003zd,Broniowski}. Within the quark
combination picture,  one can attribute it to that the rapidity
distribution of net-quarks coming from the colliding nuclei is
different from that of newborn quarks excited from vacuum just
before hadronization. It is known that the nucleus-nucleus
collisions at RHIC energies are quite transparent compared to lower
AGS and SPS energies\cite{bearden:2004}. Most of the net-quarks
after collisions are not fully stopped, still carry a fraction of
initial collision energy and are located in forward rapidity region.
Therefore the longitudinal evolution and  distribution of net-quarks
should be obviously different from that of newborn quarks. In this
paper, distinguishing the rapidity distribution of net-quarks from
that of newborn quarks, we use the quark combination model to make a
detailed study of the influence of net-quarks on the yields and
rapidity distributions of various hadrons in central Au+Au
collisions at $\sqrt{s_{NN}}= 200$ GeV.

The rapidity distribution of net-baryon in high energy
nucleus-nucleus collisions reflects the energy loss of colliding
nuclei, i.e., the energy available for hadron production
(excitation)\cite{Bearden:2003}. In fact, the effective energy for
the production of newborn quarks should be the remainder after
subtracting the energy carried by net-quarks from the total
collision energy $\sqrt{s_{NN}}$. We find in this paper that all
net-quarks are not fully converted into the net-baryon,  a fraction
of net-quarks will participate in meson combination. Therefore,  can
the rapidity distribution of measured net-baryon reflect exactly the
effective energy? In this paper, we study the rapidity densities of
the net-baryon, and compare them with the rapidity densities of
net-quarks.

In the next section we give a brief description of our quark
combination model. In section III, we calculate the yields and
rapidity distributions of identified hadrons in central Au+Au
collisions at $\sqrt{s_{NN}}= 200$ GeV, and study the influence of
net-quarks on the yields and rapidity distribution of various
hadrons. Section IV summaries our work.

\section{The quark combination model}
In this section we give a brief description of our quark combination
model. The model was first proposed for high energy $e^+e^-$ and
$pp$ collisions \cite{Xieqb:1988,Liang:1991ya,Wang:1995ch,
Zhao:1995hq,Wang:1996jy,Si:1997ux}. It has also been applied to
the multi-parton systems in high energy $e^+e^-$ annihilations
\cite{Wang:1995gx,Wang:1996pg,Wang:1999xz,Wang:2000bv}. Recently we
have extended the model to ultra-relativistic heavy ion collisions
\cite{Shao:2004cn,Yao:2006fk,shao:2006}.

The quark production from vacuum is a very sophisticated
nonperturbative process. We have developed a simple model for quark
production that is of statistical nature without dynamic details in
$e^+e^-$ collisions\cite{Xieqb:1988}. It has been found in the
PHOBOS experiments that in the energy range $\sqrt{s_{NN}}\approx
20-200$ GeV, the total multiplicity per participating nucleon pair
($\langle{N_{ch}}\rangle/\langle{N_{part}}\rangle$) in central Au+Au
collisions scales with $\sqrt{s_{NN}}$ in the same way as
$\langle{N_{ch}}\rangle$ with $\sqrt{s}$ in $e^+e^-$ collisions
\cite{Back:2003xk}. In addition, the same effect has also been
observed in $pp$ and $p\overline{p}$ data after correcting the
leading particle effect\cite{Back:2003xk}. This suggests a universal
mechanism for particle production in strongly interacting systems at
high energies\cite{Sarkisyan:2004,Sarkisyan:2006}, which is mainly
controlled by the amount of effective energy available (per
participant pair for heavy ion collisions). Based on these
experimental facts we have extended the model to nucleus nucleus
collisions at RHIC energies in Ref.{\cite{shao:2006}}. The average
number of newborn constituent quarks and antiquarks in nucleus
nucleus collisions is
\begin{equation}
\label{eq7}
\langle{N_q}\rangle=2[(\alpha^{2}+\beta E)^{1/2}-\alpha]
\langle{N_{\rm part}}/2\rangle,
\end{equation}
where
\begin{equation}
\alpha=\beta m-\frac{1}{4}.
\end{equation}
Here $E$ is effective energy for producing quarks and antiquarks
from the vacuum excitation. In nucleus nucleus collisions at RHIC
energies, the transparency leads to the decrease of the effective
energy. To obtain the effective energy $E$, we should subtract the
energy carried by net-quarks  per participant pair from the total
energy $\sqrt{s_{NN}}$. The production of strange quarks and
antiquarks, due to the heavier mass, is suppressed in high energy
collisions. Here, we use the strangeness suppression factor
$\lambda_{s}$ to denote the relative production ratio of strange
quarks to light quarks. The number of newborn light and strange
quarks/anti-quarks follow the ratio $N_u:N_d:N_s=1:1:\lambda _s$
with $\langle{N_q}\rangle=N_u+N_d+N_s$. The average quark mass is
given by $m=(2m_u+\lambda _s m_s)/(2+\lambda _s)$, where $m_u=m_d$
is the light quark mass and $m_s$ is the strange quark mass.

Our quark combination model describes the hadronization of initially
produced ground state mesons ($36-plets$) and baryons ($56-plets$).
In principle the model can also be applied to the production of
excited states \cite{Wang:1995ch}. These hadrons through combination
of constituent quarks are then allowed to decay into the final state
hadrons. We take into account the decay contributions of all
resonances of $56-plets$ baryons and $36-plets$ mesons, and cover
all available decay channels by using the decay program of PYTHIA
6.1 \cite{Sjostrand}. The main idea is to line up $N_q$ quarks and
anti-quarks in a one-dimensional order in phase space, e.g. in
rapidity, and let them combine into initial hadrons one by one
following a combination rule (see section II of Ref.
\cite{Shao:2004cn} for short description of such a rule). Of course,
we also take into account the near correlation in transverse
momentum by limiting the maximum transverse momentum difference for
quarks and antiquarks as they combine into hadrons. We note that it
is very straightforward to define the combination in one dimensional
phase space, but it is highly complicated to do it in two or three
dimensional phase space \cite{Hofmann:1999jx}. The flavor SU(3)
symmetry with strangeness suppression in the yields of initially
produced hadrons is fulfilled in the model
\cite{Xieqb:1988,Wang:1995ch}.

\section{yields and rapidity distributions of identified hadrons}
In this section, we use the quark combination model to compute the
yields and rapidity distributions of identified hadrons in central
Au+Au collisions at $\sqrt{s_{NN}}= 200 $ GeV, and study the effect
of net-quarks on various hadrons.

\begin{figure}[!htp]
\centering
\begin{minipage}[c]{.6\textwidth}
\centering \epsfig{file=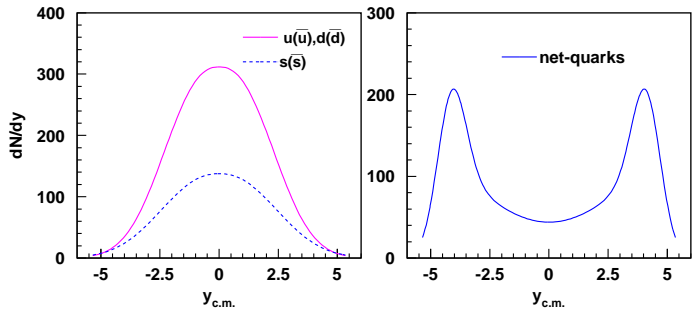,width=\linewidth}
\end{minipage}
\begin{minipage}[c]{.3\textwidth}
\centering \caption{(Color online) Rapidity spectra of newborn
quarks and antiquarks and net-quarks at hadronization in central
Au+Au collisions at $\sqrt{s_{NN}}= 200 $ GeV.}
 \label{quark}
\end{minipage}

\end{figure}

The values of parameters $\lambda _s$ and $\beta$ in Eq. (\ref{eq7})
are taken to be $\lambda _s=0.5$, $\beta=3.6\, \texttt{GeV}^{-1}$ in
accordance with Ref.\cite{shao:2006,Shao:2004cn}. The masses of
light and strange quark/antiquark are taken to be $m_{u}=0.34$ MeV
and $m_{s}=0.5$ MeV. In order to determine the effective energy $E$,
we have to get the rapidity distribution of net-quarks. Recently the
BRAHMS Collaboration have measured the net-baryon rapidity density
($dN_{(B-\overline{B}\,)}/dy$) as a function of rapidity in central
Au+Au collisions at $\sqrt{s_{NN}}=200$ GeV \cite{bearden:2004}. The
data show that the $dN_{(B-\overline{B}\,)}/dy$ in rapidity range
$-3<y<3$ are quite small compared with that in AGS and SPS energies
(see Fig.(3) in Ref.\cite{bearden:2004}). Because of baryon quantum
number conservation, the most of net-baryon should be located in
forward rapidity region $y>3$. The rapidity distribution of
net-quarks at hadronization, due to parton-hadron duality, should
also has the similar property. In this paper, we determine the
rapidity spectrum of net-quarks by fitting available net-baryon
rapidity distribution \cite{bearden:2004} in the model and
extrapolating it to the forward rapidity region. The spectrum is
plotted in Fig.\,\ref{quark}. The number of net-quarks in the entire
rapidity region should be three times of the number of participants.
One can see that only a small part of net-quarks are fully stopped
in collisions and retained in midrapidity region, the main part
still carry a fraction of initial collision energy and
concentratively appear in forward rapidity. The effective energy for
producing newborn quarks and antiquarks is written as
\begin{equation}
E=\sqrt{s_{NN}}- \frac{1}{N_{part}/\,2}\int^{y_{beam}}_{-y_{beam}}
\langle{m_T}\rangle\ cosh\,y \frac{dN_{net-quarks}}{dy} \,dy.
\label{ave}
\end{equation}
The $\langle{m_T}\rangle=\sqrt{\langle p_{T}\rangle ^2+m^2}$ is
average transverse mass of net-quarks. Since the main part of
net-quarks are located in forward rapidity region, the $\langle
p_{T}\rangle$ of net-quarks is approximately taken to be one third
of the value of net-proton at forward rapidity $y\approx
3$\cite{bearden:2004}, i.e. $0.84/3=0.28$ GeV. The effective energy
$E$ is $139.5$ GeV, and we can calculate the average newborn quark
number from Eq. (\ref{eq7}).

To calculate the hadronic rapidity spectra, we have to know the
rapidity distribution of newborn light and strange quarks just
before hadronization. As we know, the relativistic hydrodynamics
have been extensively applied to describe the evolution of hot and
dense quark matter produced in high energy heavy ion
collisions\cite{Kolb:2003dz,Satarov:2005}. In our previous work
\cite{shao:2006}, using a Gaussian-like shape rapidity distribution
for constituent quarks as a result of the Landau hydrodynamic
evolution, we have successfully described the charged-particle
pseudorapidity distributions at different centralities in Au+Au
collisions at $\sqrt{s_{NN}}=19.6,62.4,130,200$ GeV. We have found
that the rapidity distributions of pion and kaon in central Au+Au
collisions at $\sqrt{s_{NN}}= 200 $ GeV can not be simultaneously
reproduced by a single rapidity spectrum (see Fig.(6) in
Ref.\cite{shao:2006}). It suggests that the longitudinal evolution
of strange quarks, because of heavier mass, may be different from
that of newborn light quarks. In the present work, we further take
into account the difference in rapidity between strange quarks and
newborn light quarks. The rapidity distributions of newborn light
and strange quarks and antiquarks, shown in Fig.\,\ref{quark}, are
extracted from measured $\pi^{+}$ and $K^{+}$ respectively in the
model. With this input, we firstly calculate the yields and the
rapidity densities at midrapidity of various hadrons. The results
are shown in Table.I and agree with the data very well. One can see
that the yields of many hadrons, affected by net-quarks, are
different from that of the antihadrons. In the following text, we
will make a detailed study of the influence of net-quarks on
different kinds of hadrons.

\begin{table}[!htp]
\tabcolsep0.15in \setlength{\arrayrulewidth}{.5pt}
\renewcommand{\arraystretch}{1.5}
\caption{The yields and rapidity densities of identified hadrons in
central Au+Au collisions at $\sqrt{s_{NN}}= 200 $ GeV. The pion
yields are corrected for the contribution of hyperon $(\Lambda)$ and
neutral kaon $K^{0}_{s}$ decays. The $\Lambda$ yields are corrected
for feed-down from multistrange baryon weak decays. The data of
$K_{s}^{0}$ is the fit result according to its transverse momentum
spectrum given by STAR Collaboration \cite{Admas2006b}. The data of
pions and kaons are taken from \cite{Bearden:2004y,Adler:2004},
while the data of strange hyperons are taken from
\cite{Admas2006a}.}
\begin{tabular}[t]{c c c|c c }\hline \hline
\multicolumn{3}{c}{yield} \vline &\multicolumn{2}{c}{$\frac{dN}{dy}|_{y=0}$} \\ \hline
&data&model&data&model \\ \hline
$\pi^{+}$        & $1640\pm16\pm131$ & $1644$  & $286.4\pm24.2$ & 285.68   \\ \hline
$\pi^{-}$        & $1655\pm15\pm132$ & $1656$  & $281.8\pm22.8$ & 286.56   \\ \hline
$K^{+}$          & $285\pm5\pm23$    & $274$   & $48.9\pm6.3$   & 47.5     \\ \hline
$K^{-}$          & $239\pm4\pm19$    & $223$   & $45.7\pm5.2$   & 44.6     \\ \hline
$p$              &                   & 253.67  &                & 26.2     \\ \hline
$\overline{p}$   &                   & 87.73   &                & 18.82    \\ \hline
$K^{0}_{s}$      &                   & 245.28  & $45.28^{\,*}$  & 45.75    \\ \hline
$\Lambda$        &                   & 109.93  & $16.7\pm0.2\pm1.1$   & 15.51   \\ \hline
$\overline{\Lambda}$
                 &                   & 52.88   & $12.7\pm0.2\pm0.9$   & 12.05   \\ \hline
$\Sigma ^{+}$    &                   & 30.92   &                      & 4.01    \\ \hline
$\overline{\Sigma}^{\,_-}$
                 &                   & 13.64   &                      & 3.08    \\ \hline
$\mathrm{\Xi^{-}}$
                 &                   & 12.83   & $2.17\pm0.06\pm0.19$ & 2.12    \\ \hline
$\mathrm{\overline{\Xi}^{\,_+}}$
                 &                   & 7.65    & $1.83\pm0.05\pm0.2$  & 1.76    \\ \hline
$\mathrm{\Omega^{-}}$
                 &                   & 1.55    & $0.53\pm0.04\pm0.04$  & 0.31    \\ \cline{1-3} \cline{5-5}
$\mathrm{\overline{\Omega}}^{_+}$
                 &                   & 1.14    &
\raisebox{0.6ex}[0pt]{( $\mathrm{\Omega^{-}}+\mathrm{\overline{\Omega}}^{_+}$)}
                                                                      & 0.274  \\ \hline \hline
\end{tabular}
\label{yield}
\end{table}

\begin{figure}[!htp]
\centering
\begin{minipage}[c]{.45\textwidth}
\centering
\epsfig{file=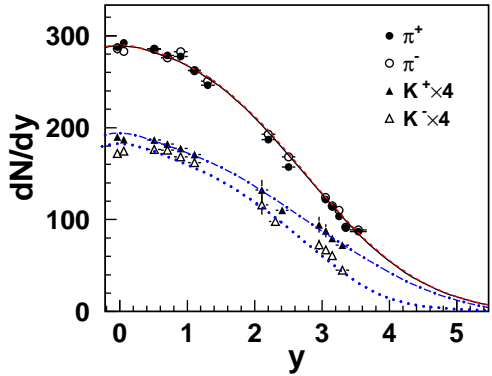,width=\linewidth}
\end{minipage}
\begin{minipage}[c]{.45\textwidth}
\centering \caption{(Color online) Pions and kaons rapidity
densities as a function of rapidity in central Au+Au collisions at
$\sqrt{s_{NN}}= 200 $ GeV. The kaons yields were multiplied by 4 for
clarity. The solid, dashed, dotted-dashed, dotted lines are our
results for $\pi^{+}$, $\pi^{-}$, $K^{+}$ and $K^{-}$ respectively.
The experimental data are given by BRAHMS Collaboration
\cite{Bearden:2004y}.} \label{pion-kaon-y}
\end{minipage}
\end{figure}

Firstly we study the influence of net-quarks on different kinds of
mesons. We divide the ground state mesons except for neutral mesons
into two types. The first type includes $K^{\pm}$,
$K^{0}(\bar{K}^{0})$, $K^{*\pm}$ and $K^{*0}(\bar{K}^{*0})$. The
constituents of these hadrons contain $u$/$d$ quark while their
antihadrons do not.  Taking the $K^{\pm}$ as an example, we compute
their rapidity densities as a function of rapidity in central Au+Au
collisions at $\sqrt{s_{NN}}= 200 $ GeV. The results are shown in
Fig.\,\ref{pion-kaon-y}. In a no net-quarks environment, e.g. in
$e^+e^-$ annihilation, the yield of $K^{+}(u\bar{s})$ would be equal
to that of $K^{-}(\bar{u}s)$. In nucleus nucleus collisions,
however, the net-quarks from colliding nuclei lead to the number of
$u$ quarks is greater than that of $\bar{u}$. The yield of
$K^{+}(u\bar{s})$ produced by direct combination would be greater
than that of $K^{-}(\bar{u}s)$. The excess of $K^{+}$ over $K^{-}$
depends on the relative quantity of net-quarks. In addition, decay
contributions from other particles $(K^{*},\phi,\Omega)$ further
increase this excess. Therefore the total rapidity densities of
$K^{+}$ are greater than that of $K^{-}$ in the full rapidity range.
The second type includes $\pi^{\pm}$ and $\rho^{\pm}$, in which both
hadrons and antihadrons have light constituent quark. We take
$\pi^{+}(u\bar{d})$ and $\pi^{-}(\bar{u}d)$ as an example, and
calculate their rapidity densities as a function of rapidity in
central Au+Au collisions at $\sqrt{s_{NN}}= 200 $ GeV. The pions
yields are collected excluding the contribution of hyperon
$(\Lambda)$ and neutral kaon $K^{0}_{s}$ decays. The effect of
net-quarks on directly produced $\pi^{+}$ and $\pi^{-}$ is embodied
in $u$ quarks and $d$ quarks respectively. The number of newborn $u$
quarks is equal to that of newborn $d$ quarks in the model. The
ratio of the number of  $u$ quarks to $d$ quarks in net-quarks is
taken to be the value in gold nucleus ($u/d=0.878$). The yield of
directly produced $\pi^{-}$($N=126$) therefore is a little greater
than that of $\pi^{+}$($N=122$). As we know, the most part of pions
comes from the decay of other hadrons and antihadrons. The yields of
these hadrons, influenced by net-quarks, are also different from
that of antihadrons. The yield of $\pi^{-}$ from hadron decay
($N=1530$) is found to be  a little more than that of $\pi^{+}$
($N=1522$). The total yield difference between $\pi^{-}$ and
$\pi^{+}$ is given as
\begin{equation}
\frac{N_{\pi^{-}}-N_{\pi^{+}}}{N_{\pi^{-}}+N_{\pi^{+}}}=\frac{1656-1644}{1656+1644}
=0.004,
\end{equation}
which is a very small value. Thus the rapidity densities of
$\pi^{+}$ and $\pi^{-}$ are  nearly equal within the  entire
rapidity region. One can see that the model  describes well the
rapidity densities ${dN}/{dy}$ of pions and kaons as a function of
rapidity in central Au+Au collisions at $\sqrt{s_{NN}}= 200 $ GeV.

\begin{figure}[!htp]
\centering
\begin{minipage}[t]{.45\textwidth}
\centering \epsfig{file=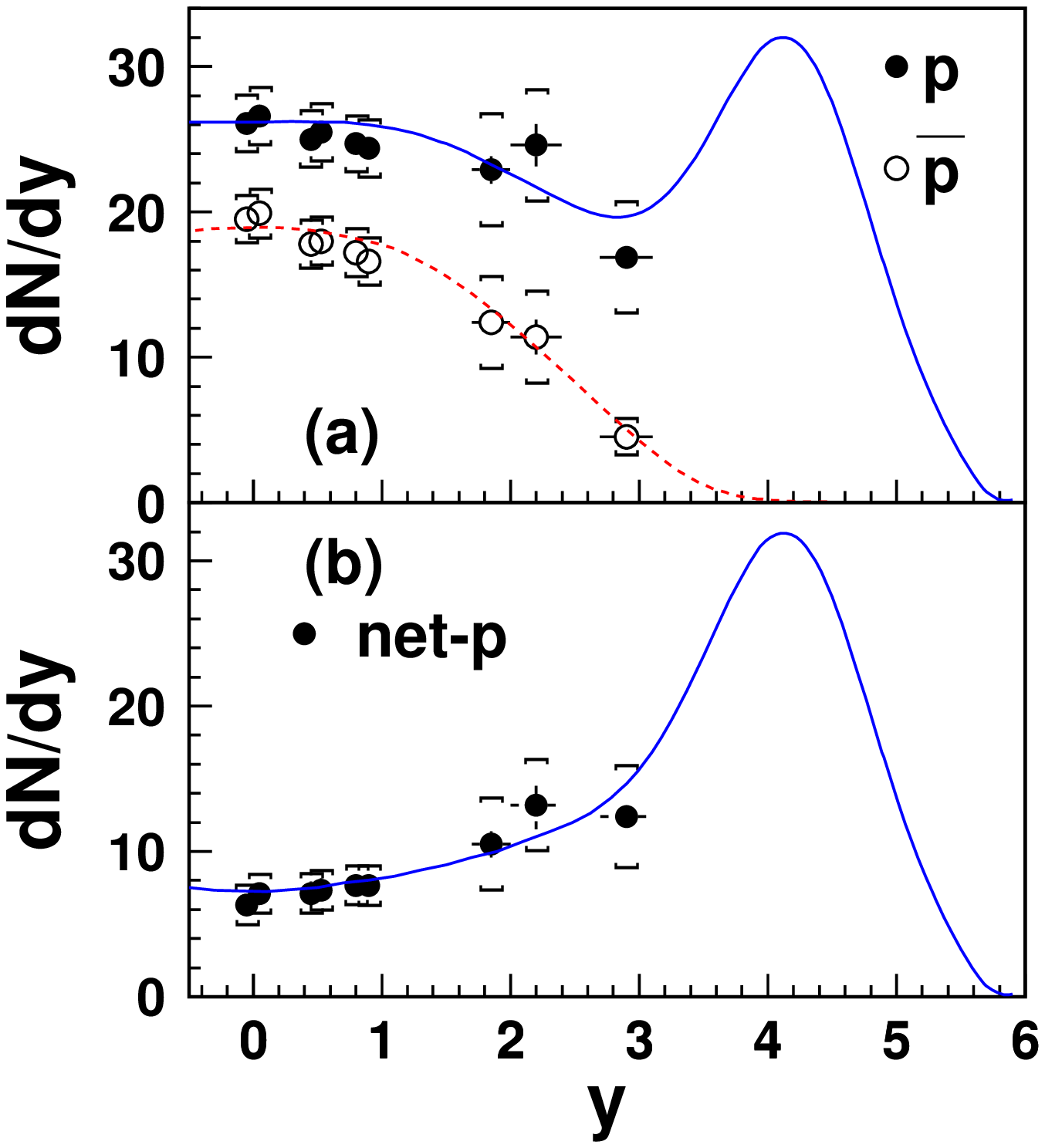,width=\linewidth}
\end{minipage}
\begin{minipage}[t]{.45\textwidth}
\centering \epsfig{file=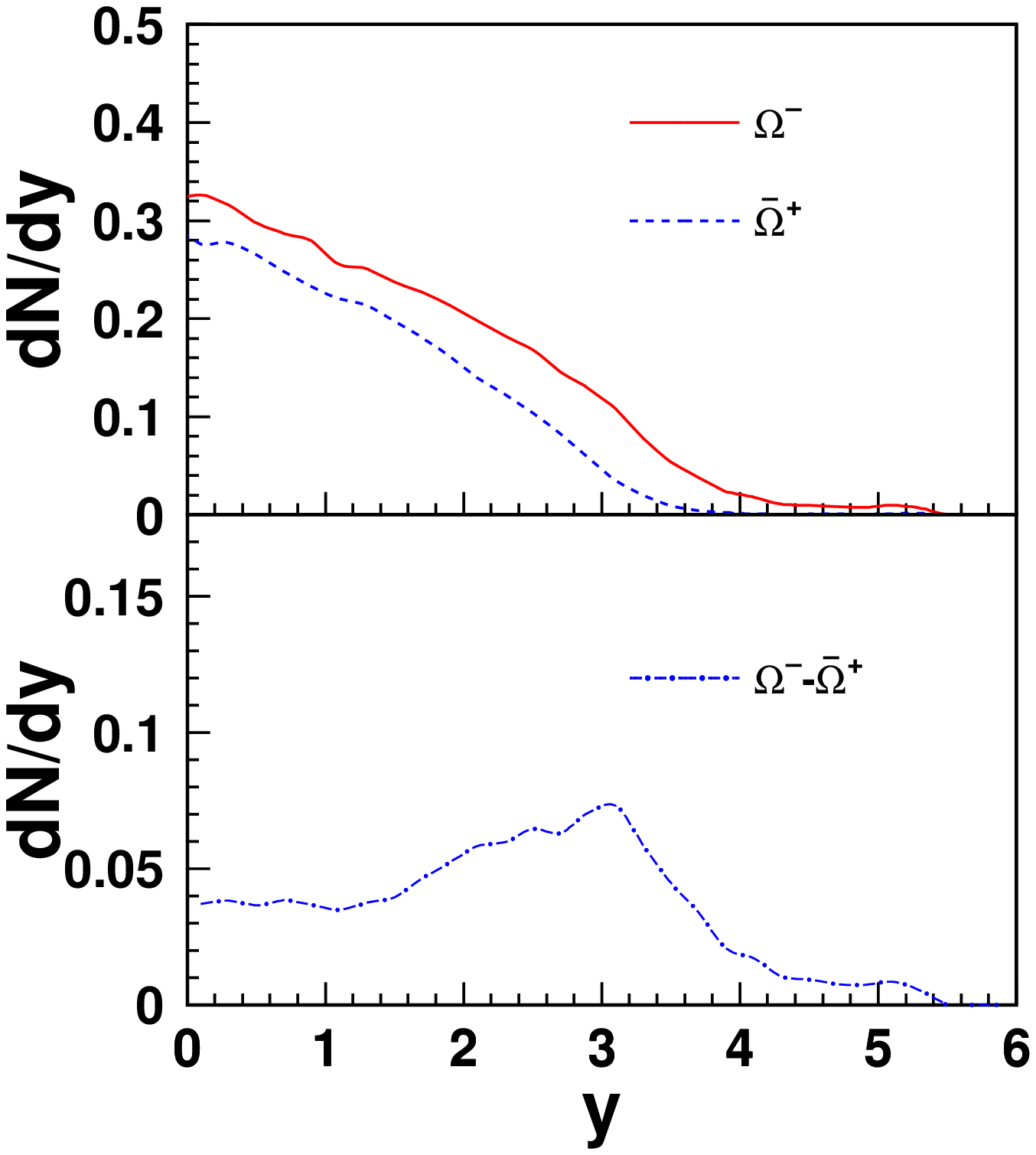,width=\linewidth}
\end{minipage}
\caption{(Color online) The rapidity densities ${dN}/{dy}$ for
proton, antiproton, and $\Omega$, $\overline{\Omega}$  as a function
of rapidity in central Au+Au collisions at $\sqrt{s_{NN}}= 200 $
GeV. The errors shown with caps include both statistical and
systematic. The experimental data are given by BRAHMS Collaboration
\cite{Bearden:2004y}.}
 \label{Fig.5}
\end{figure}

Subsequently we study the influence of net-quarks on different kinds
of baryons.  We calculate the rapidity distributions of proton,
antiproton, and net-proton ($p-\overline{p}$) in central Au+Au
collisions at $\sqrt{s_{NN}}= 200 $ GeV. The results are shown in
Fig.\,\ref{Fig.5}. No weak decay correction has been applied. The
existence of net quarks leads to the excess of directly produced
proton over antiproton in the entire rapidity region.  Taken into
account the decay contributions from hyperons such as $\Lambda, \Xi$
etc, this excess will further increase. The total excess, i.e.
net-proton ($p-\overline{p}$), at different rapidities strongly
depends on the relative rapidity densities of net-quarks. One can
see that the quark combination model can also successfully explain
the data of proton, antiproton, and net-proton in the rapidity
region $0<y<3$. For hyperons $\Omega (sss)$,
$\overline{\Omega}(\bar{s}\bar{s}\bar{s})$, their constituents do
not include the light quarks and antiquarks. On the face of it, the
yield and rapidity densities of $\Omega$ should be equal to that of
$\overline{\Omega}$. However the calculated rapidity densities of
$\Omega$, $\overline{\Omega}$ and net-Omega
($\Omega-\overline{\Omega}$), shown in Fig.\,\ref{Fig.5}, indicate
that it is not the case. Though the existence of net-quarks do not
directly affect the production of $\Omega$ and $\overline{\Omega}$,
they do affect the circumstance in which strange quarks(antiquarks)
combine into $\Omega$ ($\overline{\Omega}$). In the above paragraph,
we have pointed out that the yield and rapidity densities of $K^{+}$
are greater than that of $K^{-}$. This will lead to the number of
used $\bar{s}$ quarks is greater than that of $s$ quarks as kaons
are combined. The same situation occurs in $K^{0}(\bar{K}^{0})$,
$K^{*\pm}$ and $K^{*0}(\bar{K}^{*0})$. But the number of used
$\bar{s}$ quarks is smaller than that of $s$ quarks in strange
hypersons $\Lambda$,$\Sigma$ and $\Xi$ combination.
 The competition of these two effects causes the difference in
 the yield and rapidity densities  between
$\Omega$ and $\overline{\Omega}$ in central Au+Au collisions at
$\sqrt{s_{NN}}= 200 $ GeV.

\begin{figure}[!htp]
\centering
\begin{minipage}[c]{.45\textwidth}
\centering \epsfig{file=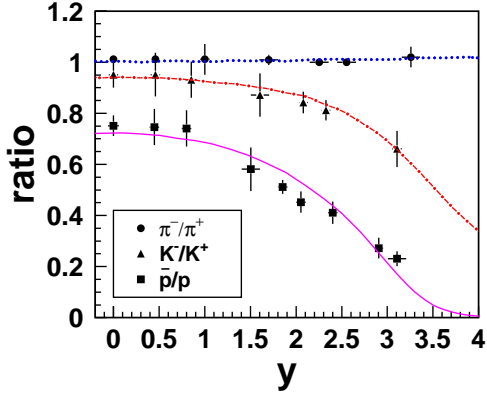,width=\linewidth}
\end{minipage}
\begin{minipage}[c]{.45\textwidth}
\centering \caption{(Color online) Antiparticle-to-particle ratios
as a function of rapidity in central Au+Au collisions at
$\sqrt{s_{NN}}= 200 $ GeV. The dotted, dotted-dashed and solid lines
are our results for $\pi^{-}/\pi^{+}$, $K^{-}/K^{+}$,
$\overline{p}/p$ respectively. The experimental data are from BRAHMS
Collaboration \cite{Bearden:2003}. } \label{pi-k-p-ratios}
\end{minipage}
\end{figure}

Rapidity dependence of the multiplicity ratios for charged
antihadron to hadron are significant indicators of the dynamics of
high energy nucleus-nucleus collisions
\cite{Herrmann:1999,Satz:2000}. We have computed the rapidity
spectra of charged pions, kaons, proton and antiproton, and it is
convenient to give the ratios of antihadron to hadron as a function
of rapidity. The results are in Fig.\,\ref{pi-k-p-ratios}. No weak
decay correction has been applied. The ratio of $\pi^{-}/\pi^{+}$ is
consistent with unity over the entire rapidity range, while the
ratios of $ K^{-}/K^{+}$ and $\overline{p}/p$ decrease with
increasing rapidity due to the influence of net quarks. In forward
rapidity region, quark matter produced in collisions are mainly
composed of net-quarks, and the newborn quarks and antiquarks only
have a quite small proportion. Both the combination probability of
quark and antiquark into meson and that of quarks into baryon are
much greater than that of antiquarks into antibaryon. Thus the
production of antibaryons in forward rapidity region becomes rather
difficult compared with baryons and mesons. The calculated rapidity
densities of antibaryons such as $\overline{p}$ and
$\overline{\Omega}$ in Fig.\,\ref{Fig.5}, are very small in forward
rapidity region. Therefore the $\overline{p}/p$ ratio decreases more
rapidly than that of $K^{-}/K^{+}$ with increasing rapidity in
forward rapidity. For example, the $K^{-}/K^{+}$ ratio reachs the
value about 0.35 at $y=4$ while the value of $\overline{p}/p$ ratio
is nearly zero at this rapidity. One can see that different rapidity
dependence of antihadron to hadron ratios can be well described by
our quark combination model.

\begin{figure}[!htp]
\centering \epsfig{file=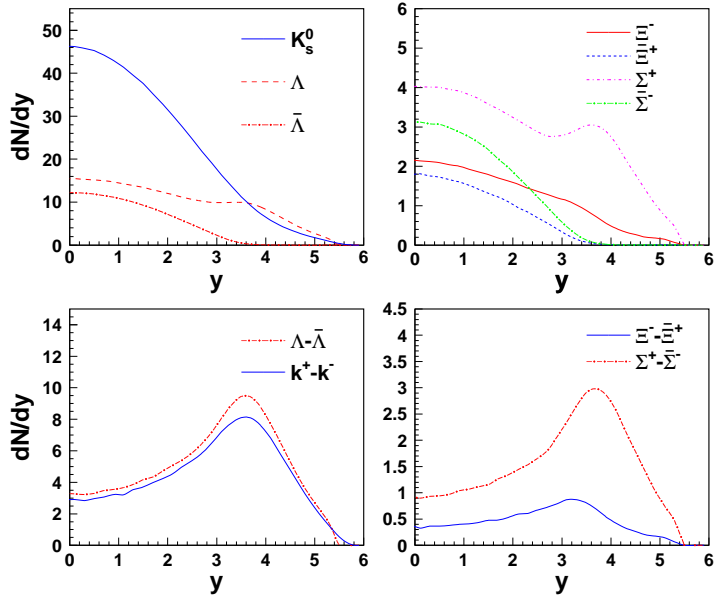,width=0.6\linewidth}
\caption{(Color online) Rapidity distributions for neutral kaon
$K^{0}_{s}$ and hyperons $\Lambda(\overline{\Lambda} )$,
$\Sigma^{+}(\overline{\Sigma}^{\,_-})$ \,$\mathrm{\Xi^{-}}$ \!
($\mathrm{\overline{\Xi}^{\,_+}}$) in central Au+Au collisions at
$\sqrt{s_{NN}}= 200 $ GeV.} \label{ystrange} \label{midrapidity}
\end{figure}

It is known that the yields and rapidity distributions of hyperons,
less affected by the hadron decay, can reflect more directly the
hadronization mechanism of the hot and dense matter produced in high
energy heavy ion collisions. We predict the rapidity spectra of
hyperons $\Lambda(\overline{\Lambda} )$,\,$\mathrm{\Xi^{-}}$ \!
($\mathrm{\overline{\Xi}^{\,_+}}$),\,$\Sigma^{+}(\overline{\Sigma}^{\,_-})$
and neutral kaon $K^{0}_{s}$ in central Au+Au collisions at
$\sqrt{s_{NN}}= 200 $ GeV. The results are shown in
Fig.\,\ref{ystrange}. The calculated rapidity densities of these
hadrons in midrapidity region agree with the data within the
experimental errors. The rapidity densities of ($K^{+}-K^{-}$)  and
the subtraction of antihyperon from hyperon are also shown in
Fig.\,\ref{ystrange}. The results clearly show the influence of
net-quarks on the rapidity densities of hyperon and antihyperon. The
rapidity distributions of these net hadrons, because of net-quarks,
all peak at forward rapidity. We observe that the peak position for
different kinds of hadrons is different. These hadrons contain the
strange quark/ antiquark constituents. Their production are affected
not only by the rapidity distribution of net-quarks but also  by the
rapidity distribution of strange quarks. The peak position is
therefore the synthetic embodiment of the two spectra. For
$\Lambda(\overline{\Lambda} )$,
$\Sigma^{+}(\overline{\Sigma}^{\,_-})$ and $K^{\pm}$, their
constituents include only one strange quark (antiquark) . They have
almost the same peak position at rapidity $y\approx 3.7$. The
hyperons $\mathrm{\Xi^{-}}$ \! ($\mathrm{\overline{\Xi}^{\,_+}}$)
and $\Omega^{-} ( \overline{\Omega}^{_-})$ have two and three
strange quarks (antiquarks) respectively. The peak positions of
rapidity distribution of ($\mathrm{\Xi^{-}}
-\mathrm{\overline{\Xi}^{\,_+}}$) and ($\Omega^{-}-
\overline{\Omega}^{_-}$), more affected by the rapidity distribution
of strange quarks and antiquarks, are different from that of
($\Lambda-\overline{\Lambda}$) etc. In addition, we find that the
rapidity densities of ($\Lambda-\overline{\Lambda}$) is slightly
greater than that of ($K^{+}-K^{-}$), but the yield of kaons is far
greater than that of hyperon $\Lambda(\overline{\Lambda} )$. We use
the relative yield difference between hadron and antihadron to
denote the influence of net-quarks
\begin{equation}
\frac{N_{K^{+}}-N_{K^{-}}}{N_{K^{+}}+N_{K^{-}}}=\frac{274-223}{274+223}=0.1\hspace{2cm}
\frac{N_{\Lambda}-N_{\overline{\Lambda}}}{N_{\Lambda}+N_{\overline{\Lambda}}}=
\frac{109.93-52.88}{109.93+52.88}=0.35
\end{equation}
One can see that the value of hyperon $\Lambda$ is much greater than
that of kaons. It suggests that the influence of net-quarks on
strange baryons is greater than strange mesons.

\begin{figure}[!htp]
\centering \epsfig{file=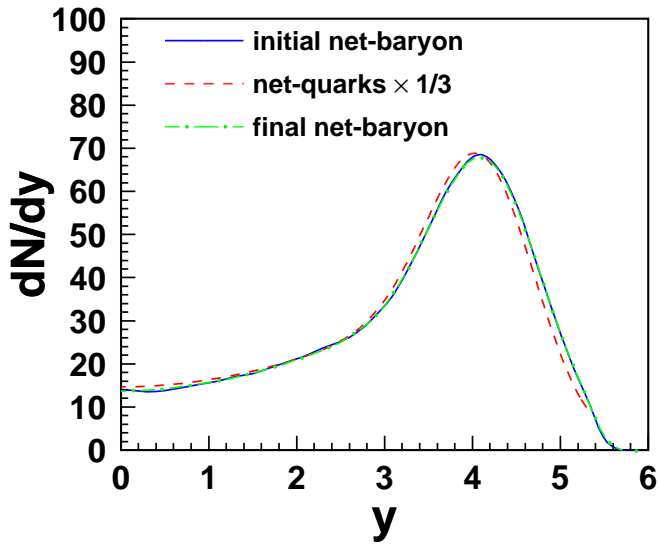,width=0.6\linewidth}
\centering \caption{(Color online) The comparison between the
rapidity distribution of net-quarks and net-baryon. The spectrum of
net-quarks is multiplied by $1/3$ } \label{compare-net}
\end{figure}

The rapidity distribution of net-baryon measured by BRAHMS
Collaboration has been used in Ref.\cite{bearden:2004} to calculate
the energy loss of colliding nuclei. However, the results shown in
Fig.\,\ref{pi-k-p-ratios} and \ref{midrapidity} indicate that not
all net-quarks are fully converted into the net-baryon, there is a
fraction of net-quarks that participate in meson combination. As we
have pointed out, the effective energy for production of newborn
quarks (hadrons) should be the remainder after deducting the energy
carried by net-quarks per participant pair from the total energy
$\sqrt{s_{NN}}$. Therefore, it is necessary to verify the
reliability and accuracy of the net-baryon-determined method. We
compute the rapidity densities of the net-baryon directly produced
by quark combination and including the hyperon decay respectively.
The results are shown in Fig.\,\ref{compare-net} and compared with
the rapidity densities of net-quarks. The rapidity distributions of
initial and final net-baryon nearly agree with that of net-quarks
except a small change of peak position around $y\thickapprox4$.
Because of the conservation of the baryon quantum number in
combination process, the net baryon quantum number carried by
net-quarks which take part in the meson combination will be
transferred to the newborn quarks nearby in rapidity. These newborn
quarks substitute for the net-quarks in mesons, and participate in
the net-baryon combination. The rapidity distribution of net-baryon
combined by these nearby newborn quarks and other net-quarks is
close to that completely combined by net-quarks. We have shown in
Ref.\cite{shao:2006} that for heavy hadrons whose masses are almost
the same as the total mass of their constituent quarks, the rapidity
distribution of hadrons nearly agrees with that of constituent
quarks. Therefore the calculated rapidity distribution of initial
net-baryon nearly agrees with that of net-quarks. Taken into account
of hyperons decay, the rapidity distribution of final net-baryon
almost has no change compared with that of initial net-baryon. The
effective energy determined by the rapidity distribution of final
net-baryon is $137.4$ GeV, which is very close to the effective
energy obtained from net-quarks. Therefore the measured net-baryon
rapidity distribution reflects exactly the energy loss of colliding
nuclei and the degree of nuclear stopping in nucleus-nucleus
collisions.

\section{Summary}
Using the quark combination model, we have systematically studied
the rapidity distributions of identified hadrons in central Au+Au
collisions at $\sqrt{s_{NN}}= 200$ GeV. It is found that while we
distinguish the rapidity spectrum of net-quarks from that of newborn
quarks, the yields and rapidity distributions of various hadrons and
antihadrons can be naturally described.

The data of net-baryon rapidity densities measured by BRAHMS
Collaboration show that the nucleus nucleus collisions at top RHIC
energy exhibit a high degree of transparency. In our work, we
extract rapidity distribution of net-quarks from the data just
before hadronization. The spectrum takes on a dual-hump shape in the
full rapidity region, which is obviously different from the
Gaussian-like distribution for newborn quarks and antiquarks. We
firstly compute the yields of various hadrons and antihadrons in the
entire rapidity region. The results are in agreement with the
available data within the experimental errors. The rapidity
distributions of identified hadrons $\pi^{\pm}$, $K^{\pm}$
$p(\overline{p})$ and net protons $(p-\overline{p})$ are reproduced.
The existence of net quarks leads to the excess of the rapidity
densities of $K^{+}$ over $K^{-}$ and that of proton over antiproton
in full rapidity range, while the difference between yields of
$\pi^{+}$ and $\pi^{-}$ resulting from net-quarks is very small
compared to the large yields of pions. The rapidity dependence of
the ratios for $\pi^{-}/\pi^{+}$, $K^{-}/K^{+}$, $\overline{p}/p$
can be well explained by our quark combination model.
 The participancy of
net-quarks indirectly affects the circumstance in which strange
quarks(antiquarks) combine into hyperon
$\mathrm{\Omega^{-}}(\mathrm{\overline{\Omega}}^{_+})$. The
calculated yield and rapidity densities of $\mathrm{\Omega^{-}}$ are
greater than that of $\mathrm{\overline{\Omega}}^{_+}$ in central
Au+Au collisions at $\sqrt{s_{NN}}= 200$ GeV. We further predict the
rapidity distribution of neutral kaon $K^{0}_{s}$ and hyperons
$\Lambda(\overline{\Lambda} )$,\,$\mathrm{\Xi^{-}}$ \!
($\mathrm{\overline{\Xi}^{\,_+}}$),\,$\Sigma^{+}(\overline{\Sigma}^{\,_-})$.
The $dN/dy$ at midrapidity of these strange hadrons are in good
agreement with the data.
 The rapidity
densities of ($\Lambda-\overline{\Lambda}$) are slightly greater
than that of ($K^{+}-K^{-}$), but the yield of kaons is far greater
than that of hyperon $\Lambda(\overline{\Lambda} )$. It suggests
that the influence of net-quarks on strange baryons is greater than
strange mesons. Finally we give the rapidity densities of the
net-baryon and compare the results with that of net-quarks. It is
found that they are nearly identical. This indicates that the
measured net-baryon rapidity distribution reflects exactly the
energy loss of colliding nuclei in nucleus-nucleus collisions.

\subsection*{ACKNOWLEDGMENTS}
The authors thank Q. Wang, Z.-T. Liang and W. Han for helpful
discussions. The work is supported in part by the National Natural
Science Foundation of China under the grant 10775089, 10475049 and
the science fund of Qufu Normal University.

\end{document}